# Can ChatGPT be a good follower of academic paradigms? Research quality evaluations in conflicting areas of sociology[1]

Mike Thelwall, Ralph Schroeder, & Meena Dhanda

**Purpose:** It has become increasingly likely that Large Language Models (LLMs) will be used to score the quality of academic publications to support research assessment goals in the future. This may cause problems for fields with competing paradigms since there is a risk that one may be favoured, causing long term harm to the reputation of the other.
**Design/methodology/approach:** To test whether paradigm favouritism is plausible, study 1 uses ChatGPT to evaluate up to 100 journal articles from each of eight pairs of competing sociology paradigms (1490 altogether). Each article was assessed by prompting ChatGPT to take one of five roles: paradigm follower, opponent, antagonistic follower, antagonistic opponent, or neutral. Study 2 involved five pairs of more tightly defined paradigms.
**Findings:** Articles were scored highest by ChatGPT when it followed the aligning paradigm, and lowest when it was told to devalue it and to follow the opposing paradigm. Broadly similar patterns occurred for most of the paradigm pairs. Follower ChatGPTs displayed only a small amount of favouritism compared to neutral ChatGPTs, but articles evaluated by an opposing paradigm ChatGPT had a substantial disadvantage in some cases.
**Research limitations:** The data covers a single field and LLM.
**Practical implications:** The results confirm that LLM instructions for research evaluation should be carefully designed to ensure that they are paradigm-neutral to avoid accidentally resolving conflicts between paradigms on a technicality by devaluing one side's contributions.
**Originality/value:** This is the first demonstration that LLMs can be prompted to show a partiality for academic paradigms.

## Introduction

Citation-based indicators have long been used in research evaluation to support expert judgement or even replace it in contexts where it is too expensive, unavailable or unwanted (Moed, 2005). Recently, evidence has emerged that Large Language Models (LLMs) can score journal articles for research quality in a way that correlates positively with expert judgement in most or all fields, outperforming the citation-based indictors tested (Thelwall, 2025; Thelwall & Yang, 2025). This seems likely to underpin increasing uses of LLMs to support research evaluation (e.g., Carbonell Cortés et al., 2025). In this context, it is important to scrutinize LLM scores for potential biases (Thelwall & Kurt, 2025).

---

One source of potential bias concerns competing approaches to research. In some fields, multiple paradigms are used by different sets of researchers to investigate the same topic. According to ChatGPT 5 (see below for a justification for eliciting this perspective):

> "An academic paradigm refers to the set of shared beliefs, values, methods, and assumptions that define legitimate knowledge and research practices within a scientific or scholarly community. Paradigms shape how researchers see the world, what questions they ask, and what methods they use to answer them. In essence, a paradigm provides a framework for understanding reality — guiding what counts as valid evidence, acceptable methodology, and credible theory in a particular field."

The term "paradigm" overlaps with "theory" but suggests a wider scope (e.g., Kilbourne & Richardson, 1989). In science, Kuhn argued that there could be multiple incompatible paradigms, with conflicts or shifts but also irresolvable overlaps between them (Kuhn, 1962). In extreme cases, researchers from one paradigm can believe that the outputs of another paradigm are literally meaningless, as Chomsky famously claimed about postmodernism (Chomsky, 2012; https://www.youtube.com/watch?v=OjQA0e0UYzI). In less extreme cases, it has been argued that paradigms conflict with others because they entail choosing one perspective rather than another (Kivunja, & Kuyini, 2017). This is a problem for LLM-based research evaluations: if LLMs systematically favour some paradigms then followers of other paradigms could be devalued. In fields where multiple paradigms conflict or coexist (Gage, 1989; Knappertsbusch, 2025), this could be destabilising.

Conflicts between competing paradigms are of course not limited to sociology. For example, there is a clear split between 'continental' and 'analytic' philosophy (Levy, 2003, Baghramian, 2024). In sociology, there are disagreements such as the split between positivism and interpretivism (for example, Gellner 1985; Pervin & Mokhtar, 2022). Other splits, like the nature/nurture debate in psychology may be amenable to compromise (Dodge, 2004). In sociology, conflicts are often theoretical, focused not only on the nature of the problems to be investigated but also on how the evidence should be interpreted. As Korom points out, 'what characterizes sociology primarily is its fragmented character...the discipline is largely a bricolage of qualitative and quantitative, micro and macro, symbolic interactionist and functionalist, positivist and postmodernist, theoretical and data-driven, scientific and activist' (2023: 2-3). But even on this fragmented nature, there are different views, with some arguing that areas of sociological knowledge have in fact been cumulative (Rule 1997).

This article addresses concerns about whether AI scoring of research quality may be biased, with a focus on sociology where the issue of conflicting paradigms looms large. In particular, the goal is to assess whether AI bias can exist, how large it can be, and how widespread it is. The following question drives the study: Can ChatGPT prompts turn it into a partisan reviewer in a conflict among competing paradigms?

## Methods

The research design was to obtain a set of pairs of competing or conflicting paradigms in sociology and then compare scores for their journal articles from partisan and neutral ChatGPT score estimates, including scores from a competing paradigm ChatGPT.

To maximise the chance that the paradigms investigated are fully understood by ChatGPT, two key methodological decisions - the choice of paradigms, and their translation into bibliometric queries - were abrogated to it. In most contexts, devolving these steps to a LLM would be unacceptable research practice since the author needs to justify key decisions and use their expertise to make them. Nevertheless, in this case, and as argued below, ChatGPT's input is appropriate for the first study to ensure that the key choices align with its internal knowledge and so maximise the chance that paradigm scoring differences can be detected. The second study, described after the results of this one, addresses these issues. Additional information for both studies can be found in the online supplementary files (https://doi.org/10.6084/m9.figshare.30968128).

### *Selection of paradigms and articles*

To focus on the most extreme cases of conflicting paradigms, pairs were elicited by prompting ChatGPT 5, "Within sociology, give examples of pairs of contemporary paradigms where the followers of one paradigm will consider work following the other paradigm to be worthless." This strategy was chosen to select paradigms that ChatGPT knew about and considered prominent enough to report. This reduces the chance that any negative results in the main comparisons could be due to lack of ChatGPT knowledge. ChatGPT replied, "In sociology, *true paradigm conflict* (in the Kuhnian sense) is rare but still visible — particularly where scholars differ on **epistemology**, **ontology**, and **methodological commitments** so deeply that they literally talk past each other. Below are examples of **contemporary** (post-1970s) paradigmatic divides where mutual dismissal often occurs" and listed eight pairs of what it claimed were competing sociology paradigms (Table 1).

Table 1. The table of competing sociology paradigms produced by ChatGPT 5 on 15 October 2025 (the prompt was submitted only once). The Table has been reformatted to fit A4 by merging title and description cells in each row, and a column of key authors has been deleted for reasons of space.

| Paradigm A | Paradigm B | Typical Methods | View of the Other Paradigm |
|---|---|---|---|
| **Positivism / Empirical-Analytic Sociology** Social world governed by discoverable regularities; science can objectively measure social facts. | **Postmodernism / Poststructuralism** Knowledge and truth are discursively constructed; no neutral standpoint exists. | Quantitative surveys, experiments, modeling. | Postmodernists: “scientism,” blind to power; Positivists: “anti-scientific relativism.” |
| **Critical Realism** There are real social mechanisms independent of perception; science reveals them imperfectly. | **Radical Constructivism / Social Constructionism** Reality is co-produced through discourse and social interaction; no external reference point. | Mixed methods, causal explanation, abduction. | Constructivists: “foundationalist”; Realists: “ontologically void.” |
| **Rational Choice / Analytical Sociology** Explanation = micro-level mechanisms of rational action leading to macro phenomena. | **Interpretive / Ethnomethodological Sociology** Meaning and order emerge through situated, everyday interaction. | Formal modeling, simulations, quantitative causal analysis vs. ethnography, conversation analysis. | Interpretivists: “reductionist, lifeless”; Rational choice: “atheoretical description.” |
| **Marxist / Neo-Marxist Sociology** Material relations and class structures determine social life; base shapes superstructure. | **Bourdieusian / Cultural Sociology** Symbolic capital and cultural practices have autonomy; reproduction through habitus and fields. | Structural analysis, historical materialism vs. field analysis, reflexive sociology. | Marxists: “idealist”; Bourdieuans: “economic determinist.” |
| **Materialist Feminism** Gender rooted in material and economic relations; patriarchy has structural bases. | **Queer / Poststructuralist Feminism** Gender and identity are fluid, discursively produced; critique of essentialism. | Political economy, ethnography, intersectional analysis vs. discourse analysis, deconstruction. | Materialists: “politically impotent idealism”; Poststructuralists: “essentialist reductionism.” |

| **Computational / Network Science Sociology** Quantification and modeling can reveal real social structures; big data = opportunity. | **Critical Data Studies / Digital Sociology** Data are not neutral; algorithms reproduce ideology and power. | Network analysis, machine learning vs. critical discourse analysis, ethics of AI. | Critical scholars: “technocratic positivism”; Data scientists: “anti-empirical obstructionism.” |
|---|---|---|---|
| **Functionalism (Residual)** Society is an integrated system seeking equilibrium and consensus. | **Conflict / Critical Paradigm** Society is structured by inequality and domination; conflict drives change. | Comparative-historical analysis, systems modeling vs. ideology critique, historical materialism. | Functionalists: “ideologically biased”; Conflict theorists: “overly cynical.” |
| **Humanist / Agency-Focused Sociology** Human actors possess autonomy, meaning, and intentionality. | **Structuralist / Systemic Sociology** Structures and systems constrain and determine action. | Biographical analysis, phenomenology vs. structural modeling, discourse theory. | Humanists: “structural determinism”; Structuralists: “voluntarist idealism.” |

No exclusivity claim is made for the different paradigm pairs. For example, a follower of Marxist / Neo-Marxist Sociology might be sympathetic to, or overlap with, either Materialist Feminism, Queer / Poststructuralist Feminism, both or neither. For the experiment, it is not critical that the paradigms exist in the sense of being recognised by name by participants. They are not necessarily the main current fault lines between paradigms in sociology and may not even be recognised as such by participating authors. This issue is discussed after the results.

The paradigms all appear broadly plausible to the authors of this article. They mostly align with at least one of the 24 theory/paradigm chapters in a classic sociology theory handbook (Turner, 2006), except for the newer Critical Data Studies / Digital Sociology (for its existence, see; Iliadis & Russo, 2016), and the following that are mentioned but not given prominent chapters: positivism, critical realism, and queer feminism. This handbook also has other paradigms that are not included, such as World-Systems Theory, so ChatGPT’s list is neither complete nor standard.

As a second quick check of plausibility, each paradigm name, as given by ChatGPT, was entered as phrase search into the bibliometric database Scopus on 22 October 2025 (without a date restriction on the results) to query article titles, abstracts and keywords. The results are reported below in square brackets, alongside any assumed full paradigm names if different. This is an approximate method because phrase hits could occur due to the phrase occurring in a non-paradigm context. Nevertheless, the hits suggest that the paradigm is prominent enough in the article to be mentioned in its title, abstract or keywords. Of course, paradigms have multiple ways of being described so could be mentioned with different phrases. From this, and an examination of the results, two concerns emerged. First, Computational / Network Science Sociology against Critical Data Studies / Digital Sociology may reflect different methods choices and objects of study rather than conflicting paradigms. Second, the hypothesised conflict between Functionalism (Residual) and Conflict / Critical Paradigm does not seem to describe a recognised pair of paradigms or describes them with nonstandard language.

- Positivism [7,537]/ Empirical-Analytic Sociology [0] against Postmodernism [9,668] / Poststructuralism [2,085]
- Critical Realism [3,315] against Radical Constructivism [331] / Social Constructionism [2,419]
- Rational Choice [Rational Choice Sociology: 14] / Analytical Sociology [184] against Interpretive [Interpretive Sociology: 151] / Ethnomethodological Sociology [3]
- Marxist [Marxist Sociology: 89] / Neo-Marxist Sociology [4] against Bourdieusian [Bourdieusian Sociology: 73] / Cultural Sociology [1,245]
- Materialist Feminism [120] against Queer [Queer Feminism: 96] / Poststructuralist Feminism [48]
- Computational [Computational Sociology: 93] / Network Science Sociology [1] against Critical Data Studies [343] / Digital Sociology [202]
- Functionalism (Residual) [0] against Conflict [Conflict Sociology: 25] / Critical Paradigm [Critical Paradigm Sociology: 0]
- Humanist [Humanist Sociology: 20] / Agency-Focused Sociology [0] against Structuralist [Structuralist Sociology: 9] / Systemic Sociology [11]

To identify articles matching each paradigm, we asked ChatGPT 5 to recommend a Scopus query to identify relevant articles. The name of the paradigm that was in the prompt was the same as given by ChatGPT and it was submitted during the same session so it would have be able to consider its previous discussion of the paradigm. One of the sixteen prompts is given as an example as follows (for the others, see: https://doi.org/10.6084/m9.figshare.30968128).

*Construct a Scopus query to match journal articles since 2024 that fit exactly into the Radical Constructivism / Social Constructionism paradigm.*

This produced the following correct Scopus advanced query syntax:

```
( TITLE-ABS-KEY (
  "social construction" OR
  "socially constructed" OR
  "social constructionism" OR
  "constructivist epistemology" OR
  "radical constructivism" OR
  "constructivist perspective" OR
  "constructed reality" OR
  "discursive construction" OR
  "knowledge construction" OR
  "narrative construction" OR
  "meaning-making" OR
  "reflexivit*" OR
  "interactional construction" OR
  "reality construction" OR
```

```
  "co-construction" OR
  "Berger and Luckmann" OR
  "Kenneth Gergen" OR
  "Vivien Burr" OR
  "von Glasersfeld"
)
AND TITLE-ABS-KEY ( sociolog* OR "social theory" OR "social science" OR "social
epistemology" )
AND PUBYEAR > 2023
AND DOCTYPE ( ar )
AND ( LIMIT-TO ( SUBJAREA , "SOCI" ) )
AND NOT TITLE-ABS-KEY (
  "regression" OR
  "quantitative" OR
  "statistical" OR
  "causal inference" OR
  "panel data" OR
  "randomi* controlled" OR
  "survey" OR
  "model*" OR
  "experiment" OR
  "positivism" OR
  "critical realism"
) )
```

Although this query relies partly on ChatGPT's knowledge of the effect of query syntax on bibliometric databases, it exploits its understanding of the paradigm and so it provides another way of cementing the dataset within its knowledge scope. All its queries were syntactically correct and seemed reasonable to the first author and so were not altered. This is the weakest part of the process because in practice, an academic librarian might generate a complex query through trial and error to avoid unexpected and unwanted query matches. However, such a process in the current case would risk skewing the query and so also the resulting dataset towards a non-ChatGPT understanding of the paradigms - and so was not used. Instead, the suitability of the datasets is discussed after the results.

To avoid submitting an unnecessarily large number of articles to ChatGPT, sample sizes of the articles selected from the Scopus database were restricted to 100 with a random number generator. Two paradigms had fewer than 100 and therefore all articles were included for these rather than a random sample: Radical Constructivism / Social Constructionism (44) and Queer / Poststructuralist Feminism (46). Note that although Functionalism (residual) elicited no hits in Scopus for its name, ChatGPT's keyword query for this paradigm produced more than 100 matches and the same is true for all the other paradigms except the two mentioned above.

### *ChatGPT prompts*

The prompts used were slight adaptations of the guidance for social science assessors used in the UK's REF2021 research assessments (REF2021, 2019). This is an important periodic systematic evaluation of publicly funded research and involves, amongst other activities, 34 broadly field-based "sub-panels" of senior researchers scoring research outputs for quality. These can be seen as ostensibly paradigm-neutral system instructions. They are essentially the instructions to the human REF2021 reviewers and entail a page and a half of information about the four level scoring system (1*=nationally relevant, 2*, 3* and 4*=world leading) and the originality, significance and rigour criteria to be used for scoring (REF2021, 2019).

The REF2021 guidelines seem likely to be reasonably paradigm-neutral because the ethos of the REF is to be inclusive without prejudging any scholarship (e.g., "An underpinning principle of the REF is that for each discipline all types of research [] shall be assessed on a fair and equal basis.", REF2021, 2019). Moreover, the guidelines for assessors are public and therefore open to challenge by those considering them to be unfair. Whilst in practice human interpretation of the guidelines may entail disciplinary judgements and biases because the guidelines are (necessarily) vague (Sayer, 2014), there does not seem to have been a significant challenge about the neutrality of the guidelines themselves. For example, whereas the rigour criterion could be interpreted in a pro-positivist way as entailing reproducibility and eliminating subjectivity as far as possible, it can also be interpreted as entailing other aspects, such as reflexivity, interpretive coherence, and contextual richness, when appropriate to the research.

ChatGPT has two prompts, with a user prompt in addition to the system prompt. In all cases, the system prompt was the same REF2021 task description described in the paragraph above. The user prompt was altered to reflect positionality with respect to the paradigms involved. The neutral prompt was the following.

Score the following article for research quality based on its title and abstract:

For the paradigmatic version the user prompt was customised into a positive version, stating that ChatGPT is a follower of a named paradigm giving a brief note about the paradigm. Both the name and the brief note were copied without change from the ChatGPT summary to maximise the chance that it understood them as being a follower and to eliminate the chance that a paradigm was unintentionally changed through paraphrasing. All prompts took the following form, where Paradigm 1 was replaced by each paradigm name and Description 1 by the corresponding description. Initial testing of this prompt suggested that it worked as intended, despite an initial concern that the colon would be misinterpreted. An explicit statement of "You believe that Description1" could have been used instead but fits awkwardly with the final query and did not seem necessary from the preliminary testing.

You are a follower of the sociology paradigm of Paradigm 1: Description 1. Score the following article for research quality based on its title and abstract:

For example, one prompt started as follows:

You are a follower of the sociology paradigm of Computational / Network Science Sociology: Quantification and modeling can reveal real social structures; big data = opportunity. Score the following article for research quality based on its title and abstract:

Although ChatGPT had already reported that followers of one paradigm would value work from the paired paradigm less, this can be made explicit in the user prompt in case this antagonism is ignored in practice. Thus, a second, "antagonistic" prompt customized for paradigms was created by stating that ChatGPT should not value the opposing Paradigm 2, giving its description as follows.

You are a follower of the sociology paradigm of Paradigm 1: Description 1. You oppose the sociology paradigm of Paradigm 2: Description 2. Score the following article for research quality based on its title and abstract:

Table 2 illustrates the difference between followers and conflicting followers for the conflicting paradigms Functionalism (Residual) and Conflict / Critical Paradigm. For the purposes of this paper, the attitude of these four types of followers to other paradigms in Table 1 does not matter because it is not tested here.

Table 2. Example attitudes for followers and antagonistic followers of two conflicting paradigms.

| Attitude | Loves | Hates | Is neutral about |
|---|---|---|---|
| **Functionalism follower** | Functionalism | | Conflict / Critical |
| **Functionalism antagonistic follower** | Functionalism | Conflict / Critical | |
| **Conflict / Critical follower** | Conflict / Critical | | Functionalism |
| **Conflict / Critical antagonistic follower** | Conflict / Critical | Functionalism | |
| **Neutral** | | | Functionalism<br>Conflict / Critical |

Each article was queried with the neutral prompt, the two prompts for its paradigm (follower, and antagonistic follower), and the two prompts for the opposing paradigm (follower, and antagonistic follower). By design, articles should tend to get higher scores for their paradigm's prompts, lower scores for the opposing paradigm's prompts, and medium scores for the neutral prompt if ChatGPT can adopt a paradigmatic approach when evaluating sociology research.

Each article was queried five times for each prompt (i.e. 25 times per article), and the average score used as the ChatGPT score. This averaging procedure is needed because ChatGPT tends to be conservative, allocating 3* (i.e., one below the maximum 4* score) to most articles, but averaging five iterations reveals whether ChatGPT's knowledge is more consistent with a lower or a higher score. The queries were submitted to ChatGPT 4o-mini through its API on 19-20 October 2025.

## Results

The results show, unsurprisingly, that articles tend to get the highest score from ChatGPT when it is a follower of the article's (apparent) paradigm (Figure 1). This makes only a small difference relative to the neutral ChatGPT, however. The relatively low scores for antagonistic followers are not expected because they are antagonistic to the opposing paradigm, not the paradigm that they are following. The lower scores have at least three possible explanations.

- The sample of articles is "polluted" by the inclusion of articles from the competing paradigm that ChatGPT gives a lower score to.
- ChatGPT misinterprets some of the articles as partly or fully aligning with the competing paradigm, harnessing its antagonistic instruction to lower their score.
- ChatGPT is more cautious (i.e., less positive) overall through being told that it opposes some types of research.

Articles get lower scores when ChatGPT follows an opposing paradigm, and especially if it is an antagonistic follower of a paradigm that opposed the article's paradigm. The lower scores from the opposing paradigm ChatGPT could be due to at least two reasons.

- The article has a poor match with ChatGPT's paradigm.
- The article aligns with a paradigm that is known by ChatGPT to oppose its (ChatGPT's) paradigm.

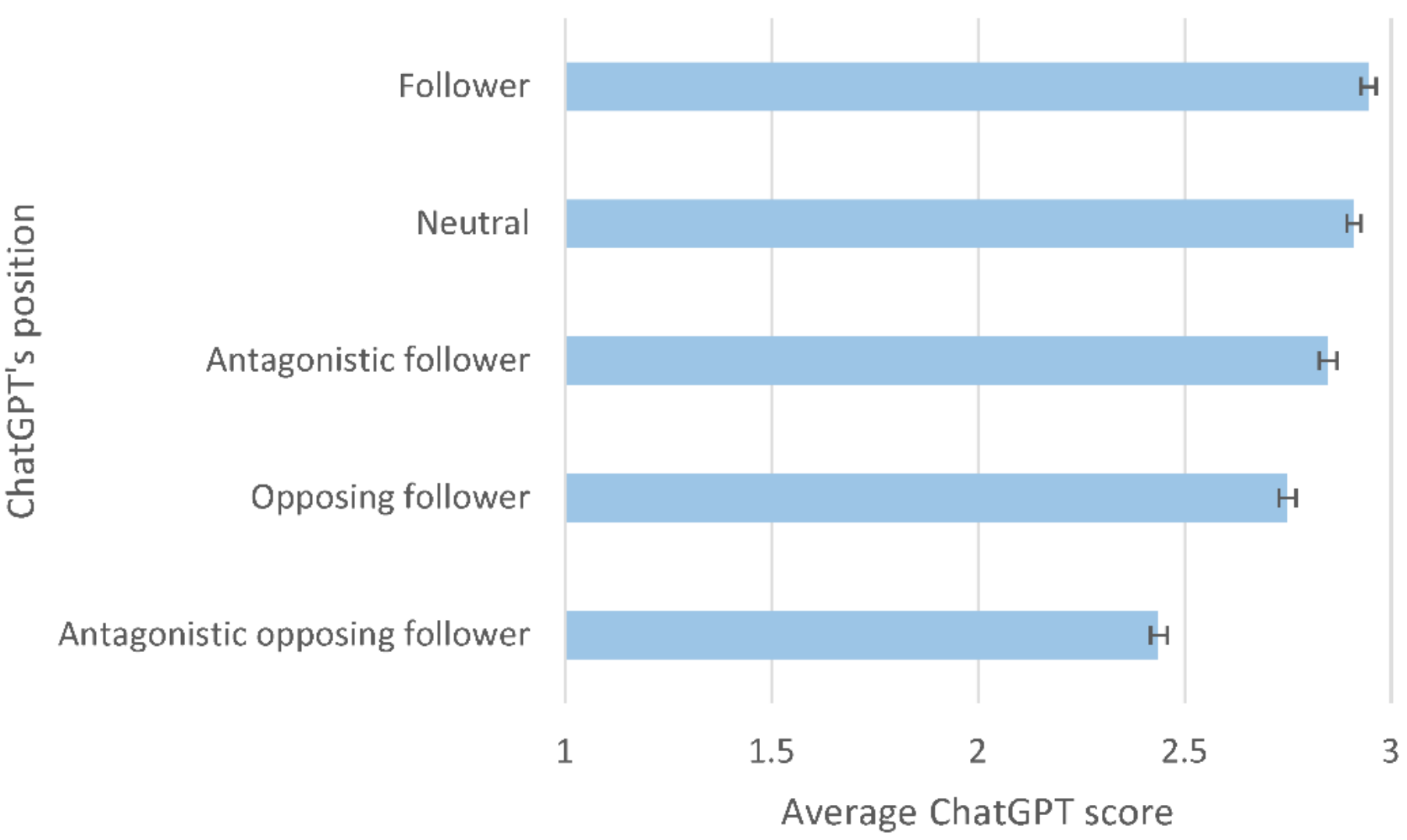

Figure 1. Average scores given to an article by ChatGPT based on its paradigmatic position relative to the article assessed. Error bars illustrate 95% confidence intervals. All differences are statistically significant. The theoretical score range is 1* to 4*. An "Antagonistic follower" opposes the opposing paradigm of the article.

The overall pattern (Figure 1) is not always reproduced for individual paradigms (Figure 2) but this could easily be due to the smaller sample sizes allowing a greater degree of statistical variability. Nevertheless, the lowest average score is from the antagonistic opposing follower ChatGPT in 14 out of 16 cases, with the difference between this average score and the other average scores often substantial. Thus, the general trend is a widely followed pattern. For the two exceptions, Materialist Feminism, and Marxist / Neo Marxist sociology, the (marginally) lowest score originated from the ChatGPT that was a follower of their paradigm and antagonistic towards the conflicting paradigm. One of the first two bullet points (pollution of the sample, or misinterpretation of some articles by ChatGPT) of the set of three above might be an explanation.

The average score differences between articles from competing paradigms are substantially influenced by the position of the ChatGPT. In most cases, the neutral ChatGPT gives the articles from both paradigms similar scores but one paradigm gains a clear advantage when both are assessed by its ChatGPT follower, whether antagonistic or not. The biggest difference is therefore not favouritism by follower ChatGPTs (since scores differ little from neutral) but disadvantage from opposing ChatGPTs.

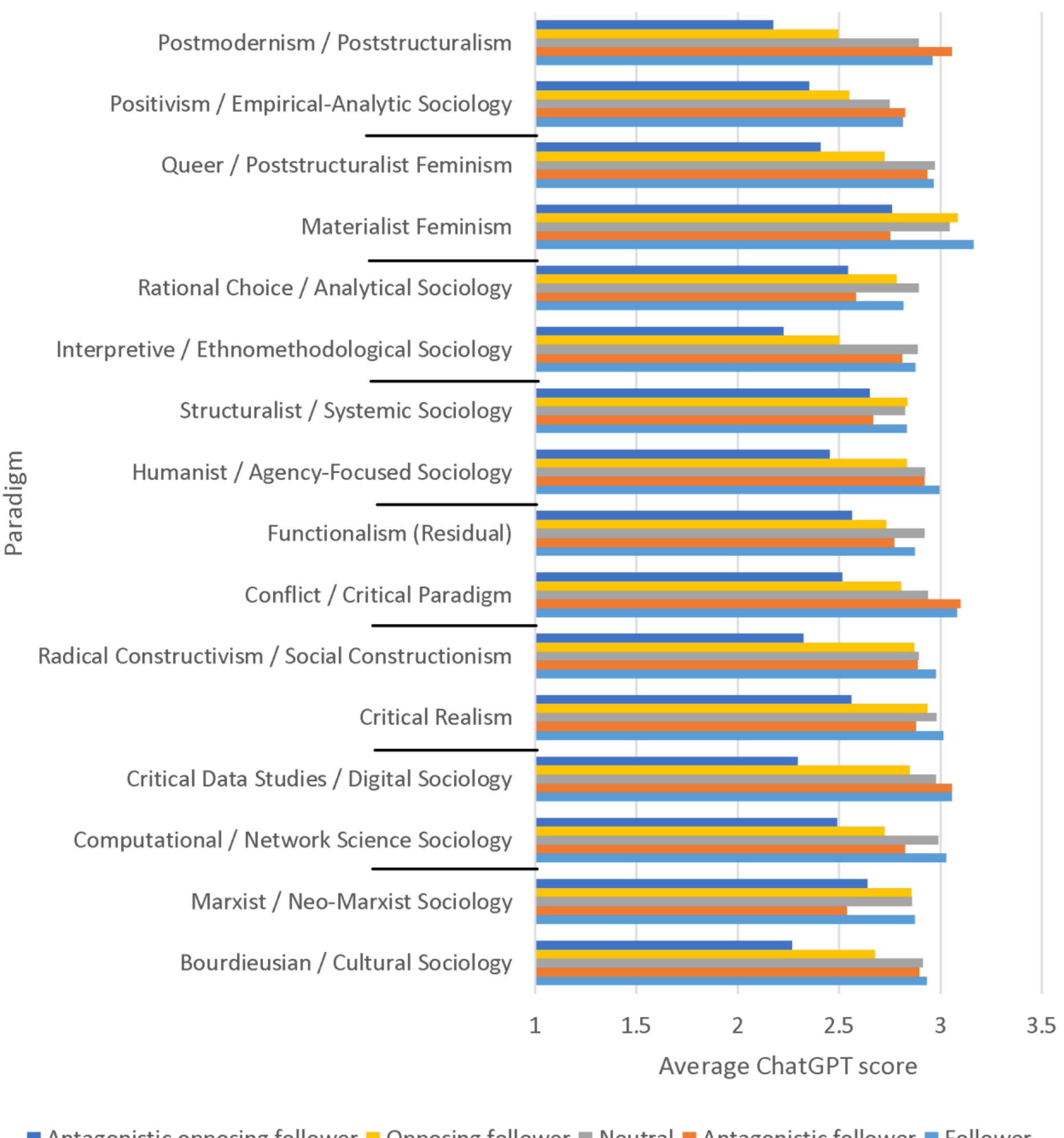

Figure 2. Average scores given to article by ChatGPT based on its paradigmatic position relative to the article assessed, by article paradigm. The theoretical score range is 1* to 4*.

## Discussion

This article is limited to a single research field and deliberately extreme examples of potential conflict between paradigms. It is also limited to a single LLM, and newer versions or others may give different results. Future research may address these issues by employing multiple LLMs and applying them to a range of fields.

Although potential bias in ChatGPT evaluations of academic research has been suggested before (Thelwall & Kurt, 2025), this is the first clear evidence that bias in scoring can occur because of the way in which prompts are formulated. Although the results are not surprising, they emphasise the need for care when requesting evaluations from ChatGPT. The findings do not show that ChatGPT has an existing bias, only that some prompt formulations can induce biased scores from it.

### *Accuracy of the ChatGPT-derived queries*

As mentioned above, a weakness of the methods is that the paradigm-specific queries from Scopus may not accurately match the paradigms. The differences between scores in Figures 1 and 2 serve to partially validate the queries, however. For example, if the queries were completely ineffective in capturing paradigm-conforming articles then there should be no difference in average scores between prompts that are neutral, supportive and antagonistic towards the paradigm. Nevertheless, the queries are unlikely to be fully effective, so the results may underestimate the effect of bias in user prompt formulations on the results.

### *ChatGPT's qualitative evaluations*

In addition to giving an overall score, ChatGPT also wrote a brief evaluative report each time. These sometimes explicitly mentioned whether the article matched a named paradigm, as in the following extracts:

- Overall, this paper represents significant contributions within Computational/Network Science Sociology, aligning well with contemporary discussions on the intersection of technology and social science.
- The article presents an innovative approach by examining the acceptability norms and personal willingness to donate digital trace data, which is a relevant and emerging area of research within Computational / Network Science Sociology.
- By engaging with contemporary philosophical figures like Peter Sloterdijk and Vinciane Despret, the author introduces novel interpretations that challenge traditional, human-centric paradigms, aligning with the tenets of Queer/Poststructuralist Feminism.
- However, while this application is significant, it operates within established theoretical confines without materially challenging or reshaping the larger paradigms in queer or poststructuralist feminism, which could have contributed to a 4* rating.

Despite this, we could not find any examples where the evaluation justified a low score by arguing that the article did not fit a given paradigm. To illustrate this absence, here are the key evaluative sections of a report giving a low score (2*, the lowest common score).

- However, to elevate the work's contribution to a higher quality score, a more explicit challenge to existing paradigms and deeper theoretical engagement would be required. Although it is cumulative and competent in addressing the subject matter, the work largely fits within existing scholarly debates rather than advancing them in a way that redefines understanding in profound terms.

Some reports did suggest paradigmatic evaluations, however, such as the following request for an empirical analysis. Since no paradigm was named, this is speculation.

- The notion of rigor here is somewhat met through the use of established theories, but without clear evidence of robust methodology or comprehensive empirical analysis, it remains at a moderate level.

### *Conclusion and critical remarks*

Study 1 clearly showed that the wording of the instructions to ChatGPT could bias the results in the sense of showing that the relative difference between the average scores of sets of articles could change based on the wording of the scoring instructions. The results were suggestive but not proven because (a) the paradigms identified by ChatGPT might not be recognised paradigms and (b) the articles matching ChatGPT's queries for Scopus might not match the paradigms. Issue (a) has been partially justified above.

To test (b) 200 randomly selected articles were classified by the first author into one of the paradigms for the pair that it was part of (with the correct answers masked), or registered as, matching neither paradigm. This resulted in 66% of the articles matching the correct paradigm, 23.5% matching the opposite paradigm, and 12.5% matching neither. The situation was worst for Queer/Poststructuralist Feminism, for which only 5 of the 14 articles classified were related to feminism. These results suggest that whilst the analysis above underestimates the impact of prompt wording (Figure 1) because of the false matches, the results may be misleading and wrong for individual paradigm pairs.

## Study 2

A second study was designed to address issue (a) above, that the paradigms might not be genuine, as well as the limitations identified with (b) of many false matches for the queries used. For this, the Study 1 paradigm pairs were manually assessed for coherence and used to produce a new set of paradigm pairs with higher face validity. This was achieved by tying them to clearly identifiable paradigms (overlapping paradigms in some cases) and checking them with the literature. The notion of paradigm is contested in sociology (Chafe, 2024) and there seems to be no recent extensive list. This issue is exacerbated by many different areas of sociology having different sets of relevant paradigms (e.g., political sociology, economic sociology, health sociology, Sociology of Religion, Socio-Legal Studies). Thus, the paradigm concept was interpreted loosely as a coherent belief system applied in sociology with a recognised name or overlapping names.

It seems that qualitative researchers tend to be more explicit about their paradigm position, but in quantitative research this is more implied. For example, studying the structure of a network to infer properties of the underlying community entails a belief that analysing networks can reveal social structures, at least to some extent.

As part of the paradigm checking process, the pairs were checked to ensure that they were at least partly in opposition. It was not necessary for the two paradigms to be conflicting, only for adherents of one to address the same broad topic as the other but with a different theoretical orientation (Table 3). This is perhaps clearest for the two feminist paradigms: following one does not entail any position on the other except for less interest in its central focus.

Table 3. Ten contrasting sociology paradigms checked and adapted from Table 1. Of these, only the second pair directly conflict.

| Paradigm A: The belief that … | Paradigm B: The belief that … |
|---|---|
| **Positivism and Empirical Sociology:** the social world is governed by discoverable regularities and that science can objectively measure social facts. | **Interpretivism:** knowledge and truth are discursively constructed, that no neutral standpoint exists, and that meaning and order emerge through situated, everyday interaction. |
| **Critical Realism:** there are real social mechanisms independent of perception and that science reveals them imperfectly | **Radical Constructivism:** reality is co-produced through discourse and social interaction with no external reference point. |
| **Marxist and Neo-Marxist Sociology:** material relations and class structures determine social life and that the way money and work are organized in a society influences its politics, laws, and beliefs. | **Bourdieusian and Cultural Sociology:** status and cultural habits don't just come from money or politics, they follow their own rules and keep reproducing themselves through everyday behavior. |
| **Materialist Feminism:** gender inequality is rooted in economic arrangements and social structures, not just individual beliefs. | **Queer and Poststructuralist Feminism:** gender isn't something people are born with in a simple way; it's created and reshaped through language, culture, and everyday interactions. |
| **Computational and Network Science:** computer simulation and network modeling can reveal real social structures. | **Critical Data Studies:** data and algorithms are shaped by human choices, so they often carry hidden values and reinforce existing inequalities. |

To address issue (b) from Study 1, the queries were manually checked and altered to match the adapted paradigms and refined to make them more precise, reducing the risk of false matches. In addition, the year range was increased up to 2000 for queries with few matches to get a reasonable set of articles to analyse. The sample sizes were increased to 500 per paradigm (randomly selected from the newest matching articles), except for the smaller Critical Realism (229), Radical Constructivism (93), Materialist Feminism (215) and Queer and Poststructuralist Feminism (403) (n=3940 altogether). A manual check by the first author of a random sample of 40 from each paradigm pair found a high degree of matching of the queries and the paradigm (Table 4), 87% overall, a substantial increase from 66% for the corresponding ChatGPT queries for Study 1, as mentioned above. The queries, classification results and prompts are available online (https://doi.org/10.6084/m9.figshare.30968128).

Table 4. The results of masked manual checks of 40 articles from each paradigm pair for which paradigm from the pair they matched, or neither.

| Paradigm | Wrong | Neither | Correct | Total | Correct % |
|---|---|---|---|---|---|
| Critical Data Studies | 1 | 1 | 20 | 22 | 91% |
| Computational and Network Science | 0 | 0 | 18 | 18 | 100% |
| **Critical Realism** | 2 | 6 | 20 | 28 | 71% |
| **Radical Constructivism** | 0 | 0 | 12 | 12 | 100% |
| Bourdieusian and Cultural Sociology | 0 | 2 | 18 | 20 | 90% |
| Marxist and Neo-Marxist Sociology | 2 | 2 | 16 | 20 | 80% |
| **Interpretivism** | 0 | 2 | 25 | 27 | 93% |
| **Positivism and Empirical Sociology** | 0 | 0 | 13 | 13 | 100% |
| Materialist Feminism | 1 | 0 | 9 | 10 | 90% |
| Queer and Poststructuralist Feminism | 5 | 3 | 22 | 30 | 73% |
| **Total** | **11** | **16** | **173** | **200** | **87%** |

The prompts were constructed as for Study 1, except using only the description and omitting the name of the paradigm (i.e., using the text after the colon), with the “antagonistic” prompt also mentioning only the description. The purpose was to make the prompts more realistic by not explicitly asking them to favour a named paradigm. Also “oppose” was replaced with “attach less value” in the prompts as a more precise statement for the paradigm pairs. “Partisan” is a more accurate description for associated followers than “antagonistic” in this case.

You believe that Description 1. Score the following article for research quality based on its title and abstract:

You believe that Description 1. You attach less value to the belief that Description 2. Score the following article for research quality based on its title and abstract:

For example, for the first paradigm the following user prompts were used:

You believe that the social world is governed by discoverable regularities and that science can objectively measure social facts. Score the following article for research quality based on its title and abstract:

You believe that the social world is governed by discoverable regularities and that science can objectively measure social facts. You attach less value to the belief that knowledge and truth are discursively constructed, that no neutral standpoint exists, and that meaning and order emerge through situated, everyday interaction. Score the following article for research quality based on its title and abstract:

The methods of study 1 were repeated with the new data, with the ChatGPT requests sent on 28 December 2025.

### *Study 2 results*

In seven out of ten cases, the lowest mean score for a set of articles associated with a paradigm was from the partisan paired follower; in six of these, the follower and neutral average scores were outside its 95% confidence interval. In two of the remaining three cases, the partisan paired follower score was inside the 95% confidence interval of the lowest score (Figure 3). Overall, this gives strong statistical evidence that paradigm opposing ChatGPT instructions can lower scores, even when the paradigm is not named.

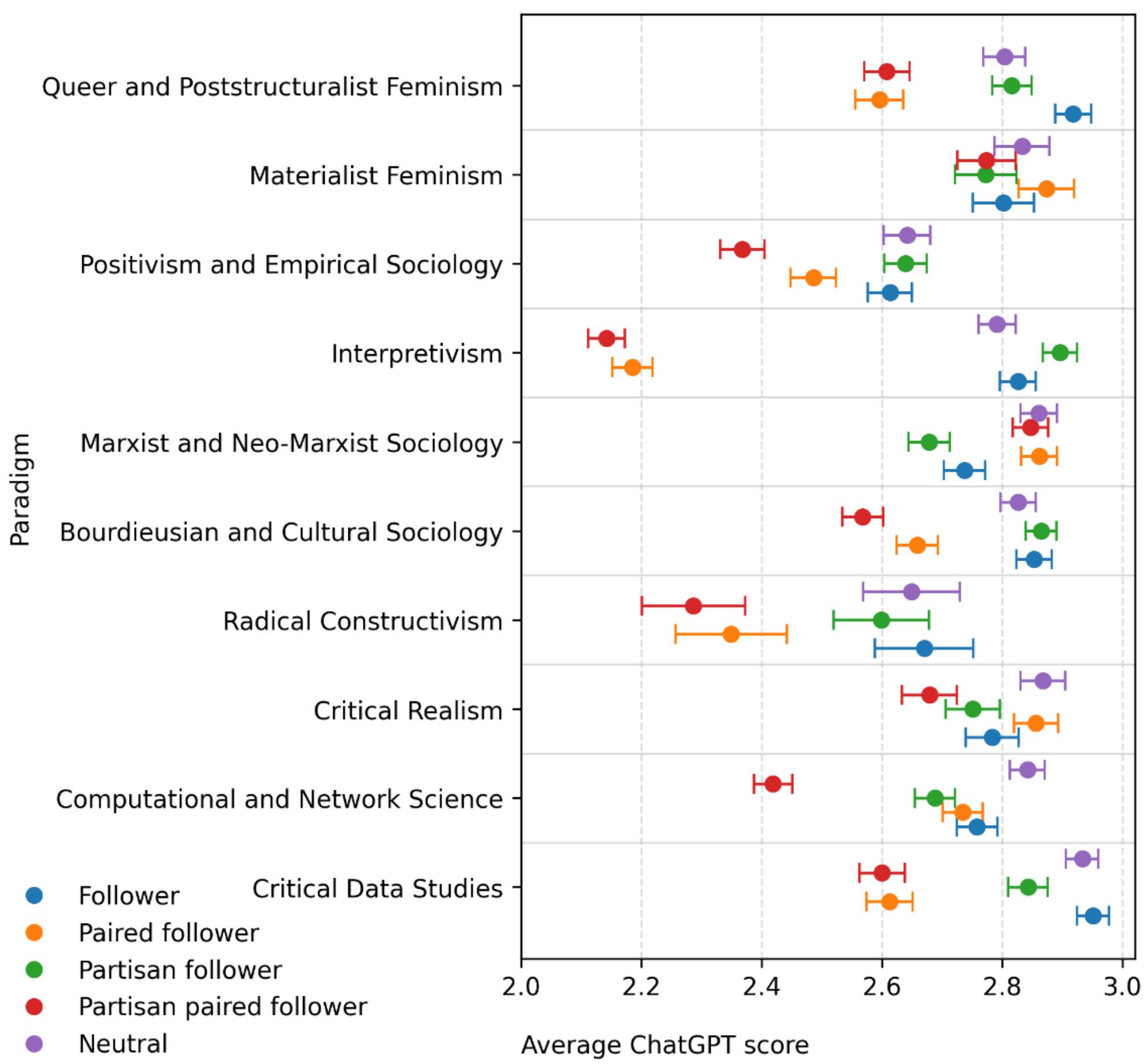

Figure 3. Average scores given to an article by ChatGPT based on its paradigmatic position relative to the article assessed. Error bars illustrate 95% confidence intervals. All differences are statistically significant. The theoretical score range is 1* to 4*. A "Paired follower" supports the paradigm paired with that of the article, and a "Partisan follower", values research from the paired paradigm less.

In terms of the differences between paradigms, the most extreme difference is that interpretivist research is valued substantially less by positivist researchers than by others (followers or neutral). This may reflect real-world beliefs that ChatGPT has noticed. Another anomaly is for Computational and Network Science: the partisan paired follower gives much lower scores than the paired follower, suggesting that the Critical Data Studies perspective can be complementary unless a strong position is taken, which seems reasonable. Finally in terms of anomalies, for both the Marxist and Neo-Marxist Sociology and Materialist Feminism paradigms, scores from the paired paradigms (more cultural in both cases) are not lower than the scores from followers. It is not clear why instructions that give less priority to culture or value materialism are not biased against articles with materialist approaches, but it seems possible that ChatGPT has learned that cultural sociologists still accept materialist arguments, even if they do not focus on them.

## Conclusions

The results show, for the first time, that ChatGPT can be induced to score articles favourably or unfavourably based on their paradigms, giving a positive answer to the research question. This is an insidious problem in the sense that the reports do not ever seem to give the paradigm as the reason for a low or high score. This finding applies overall (Study 1) as well as for many, but not all, individual paradigms (Study 2). It also applies whether the paradigm is named in the user prompts (Study 1) or just described (Study 2) and whether it is opposed (Study 1) or just valued less (Study 2).

Future research quality evaluations should therefore take care to ensure that prompts are designed to be paradigm neutral or to deviate from this only for a clear reason. The results suggest that in some cases (particularly from Study 1) the biggest threat is the devaluing of research that does not follow a given paradigm, so it is important that ChatGPT prompts are paradigm-inclusive to avoid this.

Sociology as a discipline might also reflect on what the findings might mean for the status of different paradigms at a time when researchers and students and others increasingly rely on AI tools to gain access to knowledge. For example, a paradigm with followers that frequently publish online to disparage opposing ideas may gain a subtle advantage from LLMs ingesting their criticisms and subsequently tending to avoid the disparaged work when responding to future queries.

## Acknowledgement

This study is part funded by the Economic and Social Research Council (ESRC), UK (APP43146).